\documentclass[5p,times]{elsarticle}

\usepackage{listings} 
\usepackage{makeidx}
\usepackage{algorithm,algpseudocode}
\usepackage{float}
\usepackage{amsmath}
\usepackage{rotating,todonotes, xspace}
\usepackage{multirow}
\usepackage{hyperref}
\usepackage{xspace}
\usepackage{paralist}
\usepackage{booktabs}
\usepackage{cleveref}
\usepackage{todonotes}
\usepackage{caption}
\usepackage{subcaption}                    
\usepackage{varwidth}
\usepackage[normalem]{ulem}
\usepackage{color, colortbl}
\usepackage{tabularx}
\usepackage{graphicx}
\usepackage[raggedrightboxes]{ragged2e}

\begin{document}
\begin{frontmatter}

\title{Integrating Quantum Computing Resources into Scientific HPC Ecosystems}

\author[nccs]{Thomas Beck\corref{cor1}}
\ead{becktl@ornl.gov}

\author[nccs]{Alessandro Baroni}
\ead{baronia@ornl.gov}

\author[cse]{Ryan Bennink}
\ead{benninkrs@ornl.gov}

\author[cse]{Gilles Buchs}
\ead{buchsg@ornl.gov}

\author[nccs]{Eduardo Antonio Coello P\'erez}
\ead{coellopereea@ornl.gov}

\author[nccs]{Markus Eisenbach}
\ead{eisenbachm@ornl.gov}

\author[nccs]{Rafael Ferreira da Silva}
\ead{silvarf@ornl.gov}

\author[nccs]{Muralikrishnan Gopalakrishnan Meena}
\ead{gopalakrishm@ornl.gov}

\author[nccs]{Kalyan Gottiparthi}
\ead{gottiparthik@ornl.gov}

\author[nccs]{Peter Groszkowski}
\ead{groszkowskip@ornl.gov}

\author[qsc]{Travis S.~Humble}
\ead{humblets@ornl.gov}

\author[nccs]{Ryan Landfield}
\ead{landfieldre@ornl.gov}

\author[nccs]{Ketan Maheshwari}
\ead{maheshwarikc@ornl.gov}

\author[nccs]{Sarp Oral}
\ead{oralhs@ornl.gov}

\author[nccs]{Michael A. Sandoval}
\ead{sandovalma@ornl.gov}

\author[nccs]{Amir Shehata}
\ead{shehataa@ornl.gov}

\author[nccs]{In-Saeng Suh}
\ead{suhi@ornl.gov}

\author[nccs]{Christopher Zimmer}
\ead{zimmercj@ornl.gov}

\address[nccs]{
    National Center for Computational Sciences, 
    Oak Ridge National Laboratory,
    Oak Ridge, TN, USA}

\address[cse]{
    Computational Sciences and Engineering, 
    Oak Ridge National Laboratory,
    Oak Ridge, TN, USA}

\address[qsc]{
    Quantum Science Center, 
    Oak Ridge National Laboratory,
    Oak Ridge, TN, USA}

\cortext[cor2]{\scriptsize This manuscript has been authored by UT-Battelle, LLC, under contract DE-AC05-00OR22725 with the US Department of Energy (DOE). The publisher acknowledges the US government license to provide public access under the DOE Public Access Plan (\url{http://energy.gov/downloads/doe-public-access-plan}).}

\begin{abstract}
Quantum Computing (QC) offers significant potential to enhance scientific discovery in fields such as quantum chemistry, optimization, and artificial intelligence. Yet QC faces challenges due to the noisy intermediate-scale quantum era's inherent external noise issues. This paper discusses the integration of QC as a computational accelerator within classical scientific high-performance computing (HPC) systems. By leveraging a broad spectrum of simulators and hardware technologies, we propose a hardware-agnostic framework for augmenting classical HPC with QC capabilities. Drawing on the HPC expertise of the Oak Ridge National Laboratory (ORNL) and the HPC lifecycle management of the Department of Energy (DOE), our approach focuses on the strategic incorporation of QC capabilities and acceleration into existing scientific HPC workflows. This includes detailed analyses, benchmarks, and code optimization driven by the needs of the DOE and ORNL missions. Our comprehensive framework integrates hardware, software, workflows, and user interfaces to foster a synergistic environment for quantum and classical computing research. This paper outlines plans to unlock new computational possibilities, driving forward scientific inquiry and innovation in a wide array of research domains.
\end{abstract}

\begin{keyword}
Quantum Computing, High-Performance Computing, System Integration, Quantum Algorithms, Quantum Applications. 
\end{keyword}

\end{frontmatter}

\section{Introduction}
\label{sec:introduction}

Quantum Computing (QC) holds great promise to accelerate discovery in multiple fields, including quantum chemistry, materials, optimization, national security, artificial intelligence, and the health sciences.
Any Quantum Algorithm (QA) with the potential to achieve a speedup over classical analogues relies on the coherent manipulation of quantum bits (qubits)~\cite{Nielsen&Chuang}; unfortunately, in the current Noisy Intermediate-Scale Quantum (NISQ) era, external noise tends to degrade quantum coherence before the full power of the quantum calculation can be exploited~\cite{Preskill2018quantumcomputingin}.  While advanced methods for error correction and noise mitigation are currently under active development~\cite{kim2023evidence}, it is clear that maintaining the delicate coherent state of the many-qubit system is a major scientific and technological challenge~\cite{QEC_review_2012,QEC_2023,dasilva2024demonstration,QuEra_logical_2024}.The search for quantum computers capable of exploiting the speedups offered by these QAs is evidenced by the emergence of diverse hardware technologies, including superconducting~\cite{superconducting_2019,superconducting_2020}, trapped ion~\cite{trapped_ions_2019}, optical/photonic~\cite{photonic_review_short_2019,photonic_review_long_2018}, topological~\cite{Topological_2008,Topological_2022}, quantum dot~\cite{Quantum_Dots_RMP_2023}, nitrogen vacancy centers in diamond~\cite{NV_2021}, and neutral atom qubits~\cite{Neutral_2020}. The activity focused on the development of functional qubits is clearly in a state of rapid flux. 

Due to the existence of powerful classical high performance computing (HPC) resources that can perform many compute tasks rapidly and efficiently, it seems sensible at this stage to view the Quantum Processing Unit (QPU) as an accelerator that speeds up certain demanding, exponential-scaling calculations in a scientific code. Other tasks can then be left to the classical computer, analogous to the CPU/GPU heterogeneous structure seen in modern leadership-class machines. Therefore, we are developing a broad framework for the integration of QC into HPC ecosystems, with a long-term goal of agnostic design relative to detailed hardware specifications. We also acknowledge that, practically speaking, we may have to flexibly address specific hardware features in our framework, considering the current level of development. We broadly include quantum simulators\footnote{We point out that no universally accepted definitions of a quantum \textit{emulator} or \textit{simulator} have been agreed upon by the research community, and thus, throughout this work, we will use \textit{simulator} to mean either of those.} that operate on classical hardware as part of the present-day quantum toolbox for integration development.  

The Oak Ridge Leadership Computing Facility (OLCF) at ORNL has deep expertise in standing up world-leading supercomputers that utilize GPU accelerators (Titan, Summit, and Frontier). The DOE project design and management process for the supercomputer lifecycle is thorough and mature, and we plan to use that process as a guiding principle. The steps include (a)~mission need, (b)~alternatives analysis, (c)~benchmarks, (d)~requirements gathering and request for proposals, (e)~procurement, (f)~installation, (g)~acceptance testing, (h)~optimization of a broad class of codes, and (i)~transition to operations.  We, of course, recognize that in classical HPC, the GPU accelerators are a more mature technology than current QC hardware, so adaptation to that reality will be a necessity (as noted above).  At the current stage, we will focus mainly on the first four steps (a)-(d), with added focus on (h). The present effort aimed at a general framework for incorporating quantum devices can be viewed as preparatory work, including testbed hardware evaluations, to enable full execution of the above design/build process when mature hardware options are available.  

For the mission needs and alternatives analysis, we will develop priorities for science driver applications that are best suited to the mission needs of DOE and ORNL. These drivers include the energy, earth, materials, and computational sciences, and cutting-edge supercomputer developmental research.  As part of the process, we will provide a mapping of the merits and challenges for each hardware option in relation to the various domain science applications.  This mapping will be helpful in assessing the best pairings that will guide future procurements.  

{\color{black} The private sector has dominated the development of novel quantum hardware, and it likely will continue to do so. It is the aim of this paper to develop a strategy to seamlessly integrate that hardware into our leading HPC systems in order to optimize its scientific impact.} 

\begin{figure}[!ht]
    \centering
    \includegraphics[width=\linewidth]{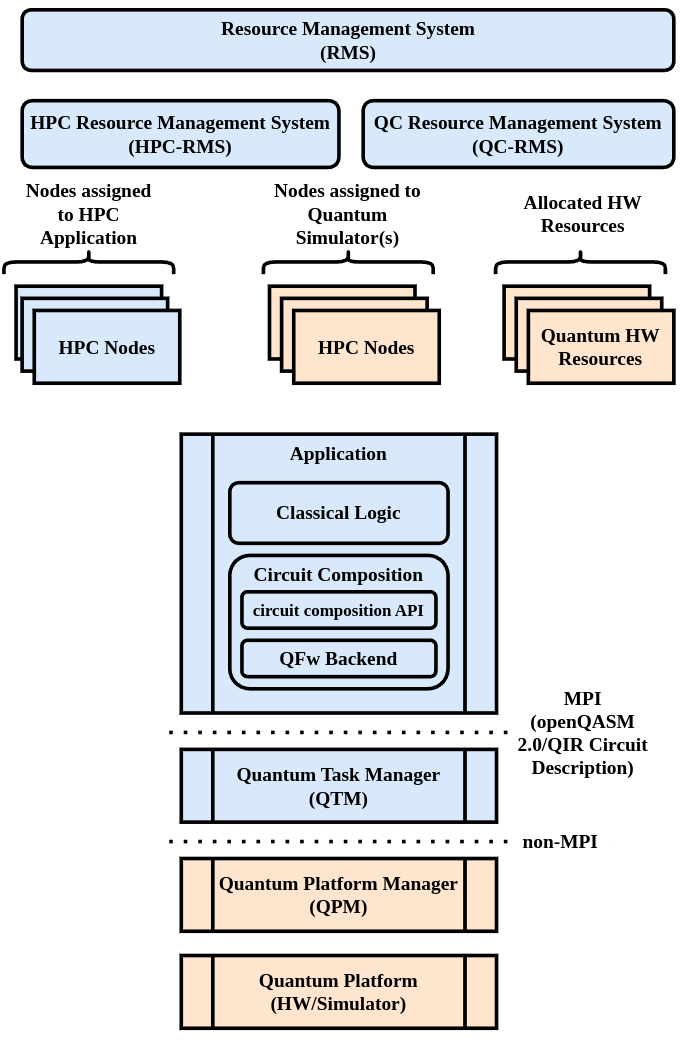}
    \caption{{\color{black}
    The QC/HPC Integration Framework uses a layered approach, with a quantum-aware resource management system reserving resources. Applications run on these resources and communicate quantum operations via MPI through the Quantum Task Manager, which can modify tasks, such as by circuit cutting. The Quantum Platform Manager then executes the prepared tasks on the quantum platform.
    Blue boxes represent classical resources, while orange boxes denote quantum resources.}}
    \label{fig:hpc_qc_architecture}
\end{figure}

The paper is organized as follows. We first survey the current state-of-the-art in QC/HPC integration and then discuss what we plan to achieve with the proposed work.  This is followed by overview discussions of hardware, users, training, applications, and the software ecosystem.  {\color{black} Following the background/motivation discussions, we then present the major focus of the paper: a detailed framework for the integration that outlines the integration space, usage patterns, and integration models, which is followed by a summary overview of the integration effort. The necessary workflows for creating a highly functional end-to-end integration strategy are discussed. Figure~\ref{fig:hpc_qc_architecture} provides a preliminary overview of the framework’s general organization. Detailed information will be presented in Section~\ref{sec:integrationFramework}. We close by presenting a summary and an outlook for the future.}

\section{Current State of the Art}

Focused activity on the development of QC is occurring in the US, Europe, Japan, Australia, Canada, China, and Russia.  Significant investments are being made in these regions due to the potentially high impact of quantum technology on basic research, the economy, healthcare, and national security~\cite{mckinseyqcreport}.  

In the US, the private-tech industry is driving the development of hardware advances, while the private sector, national laboratories, and academia are actively developing the theory and software necessary to optimize utilization {\color{black} and integration} of the quantum architectures.  At OLCF/ORNL and through collaborative agreements with QC vendors, remote access to quantum hardware is made possible through the cloud via user programs such as the Quantum Computing User Program (QCUP). This groundbreaking program facilitates access of researchers to 
superconducting 
(IBM and Rigetti) and ion trap 
(Quantinuum and IonQ) quantum computers. The QCUP  is the leading quantum user program of its kind with 80 projects and 271 users as of the end of 2023~\cite{OA2023}, with usage steadily increasing in 2024. 
While the QCUP program is a major step towards the exploration of new ideas in testing and leveraging quantum computing, this cloud-based integration approach is limited by latencies due to data transfer over long distances, dual queuing systems at both the HPC and quantum hardware sites, algorithmic and workflow constraints on passing data and codes to/from the quantum hardware, and compiling the codes. {\color{black} With the emergence of on-premisis quantum devices, developing a framework that manages the workflows for the integration of the classical and quantum compute elements is a prime research focus with significant challenges. As discussed below, such efforts are underway at multiple centers around the world~\cite{schulz2022accelerating}.} 

Within the European Union, the joint undertaking EuroHPC is coordinating activities in advanced computing research and infrastructure. In 2022, EuroHPC identified six HPC centers as hosts for future quantum computers in the EU: IT4Innovations in Czechia, Leibniz Rechenzentrum (LRZ) in Germany, Centro Nacional de Supercomputaci\'on in Spain, Grand \'Equipement National de Calcul Intensif in France, Consorzio Interuniversitario del Nord est Italiano Per il Calcolo Automatico in Italy, and Pozna\'nskie Centrum Superkomputerowo-Sieciowe in Poland. As of 2024, multiple sites are evaluating vendors for the installation of quantum computing systems while others are engaged in joint development of a QC/HPC framework. Additionally, in Europe, several countries are engaged in large-scale QC/HPC integration activities including Forschungszentrum J\"{u}lich and LRZ in Germany, and VTT Technical Research Centre in Finland. Alongside individual research efforts, these developments have focused on preparing optimized software stacks and workflows that will harness the power of various quantum technologies. 
In Japan, Rikagaku Kenky\=usho has initiated focused efforts to integrate QC with HPC through partnerships with commercial vendors as well as in-house development of quantum hardware.  
The United Kingdom (UK) has established the National Quantum Computing Centre under the auspices of the UK Research and Innovation public body to further the development of quantum computing. This includes recent investments in a bevy of commercial vendors for testing and evaluating quantum computing architectures.
Related efforts are underway at the Pawsey Supercomputing Research Center in Australia. 

On the theoretical and computational sides, extensive research around the world is being directed at a wide range of fundamental and applied algorithms in both computer science and the traditional HPC areas (simulation of quantum many-body systems, solution of partial differential equations, machine learning, and optimization, among others). A recent review and white paper led by IBM provides a detailed overview of progress in quantum computing for the materials sciences~\cite{alexeev2023quantum}. {\color{black} A sampling of other specific quantum computing applications includes protein structure prediction~\cite{doga2024perspective}, quantum dynamical simulations~\cite{fedorov2021ab}, condensed matter physics~\cite{chertkov2023characterizing}, quantum chemistry~\cite{cao2019quantum}, and fundamental studies of quantum error correction~\cite{babbush2021focus}.}  A thorough review of the status of quantum algorithms has appeared recently~\cite{dalzell2023quantum}. 

To summarize, there is vigorous worldwide research activity in quantum computing, including extensive work on QC/HPC integration. It is generally agreed that QC will find, for the foreseeable future, its greatest use as an accelerator for several harsh-scaling or otherwise intractable classical algorithms that are featured in various scientific applications. Thus, a great deal of effort is being directed toward incorporating fully functional error-tolerant quantum devices as they appear.

\section{Goals and Activities}

At ORNL, we have accumulated extensive experience in accessing quantum hardware, coordinating with HPC systems, and transferring data from classical to and from quantum hardware.  We are thus already in a position to adapt and handle physical quantum hardware onsite as part of a testbed model, similar to other HPC centers around the world. So far, these developments have been hardware and software specific. In addition to our goal of developing an integration framework to facilitate research at ORNL (Section~\ref{sec:integrationFramework}), we view the present effort as enabling the basic and applied research communities around the world to engage in QC research through the user program. A principal aim is to broaden and generalize our capabilities by developing a seamless QC/HPC integration framework. 

We have organized our QC/HPC integration work plan into five categories: hardware, users, applications, software, and framework development including workflows. Following this brief overview of the goals and activities, more detailed discussions appear in the following sections.  

A major focus is to deploy our integration framework as a set of tools to facilitate the stages in the HPC project lifecycle management process listed in Section~\ref{sec:introduction}. The framework can then be viewed as a testbed ecosystem that provides the infrastructure for analyzing quantum devices and software development activities. This work will then feed into the alternatives analysis, benchmarking, requirements, and code optimization activities in the project.  

The impact of QC/HPC on the DOE workload has been sufficiently explored in several workshops and resulting documents~\cite{dodge_ge_2024, osti_2001045, osti_1986455, osti_1470992, Moore2017_BES, osti_1616258, osti_1194404}.
Here, we mention that there are two potential major benefits to the successful incorporation of QC in scientific computing: energy efficiency and (not unrelated) algorithmic superiority. Both of these factors will assume ever-increasing importance in the worldwide computing landscape.  

For the alternatives analysis, we will perform a careful study that analyzes the attributes of each architecture (superconducting circuits, trapped ions, electron and/or nuclear spins, optical/photonic, topological, etc.).  This activity will include analysis of compute speed, coherence time, data upload and offload, noise, error rates and correction strategies, and other factors  (such as infrastructure and energy consumption) that influence the integration into an HPC system.  
Coupled with the hardware analysis, we will prioritize a list of applications driven by the DOE/ORNL mission (see Section~\ref{sec:introduction}). The list will include major applications in materials simulation and quantum chemistry~\cite{mcardle2020quantum}, optimization~\cite{lykov2023fast} (e.g., electric grid modeling), Quantum Machine Learning (QML)~\cite{sajjan2022quantum}, Computational Fluid Dynamics (CFD) modeling~\cite{jaksch2023variational, li2023potential}, other Partial Differential Equation (PDE) problems, and national security.  

A crucial step will be to then map the domain problems with their given characteristics to the hardware capabilities. For example, a low-latency configuration would be required for a quantum simulation that requires many solutions (up to millions) to obtain the Born-Oppenheimer ground-state electronic potential surface on which the nuclei move~\cite{fedorov2021ab}. This detailed mapping will help lay the groundwork for the decision process for procurement of one or more quantum devices.  We will answer questions such as: Is one general-purpose quantum computer sufficient to cover the needs arising from most applications, or are two or more accelerators necessary for algorithmically different tasks?  While the goal is an application-agnostic quantum device, invariably, some processor architectures will excel for specific computing operations.   

Since the strategies that link hardware to application algorithms and then integrate the HPC and quantum aspects of the problem involve an optimization in a very high dimensional space, we will investigate the use of (classical) Machine Learning (ML) methods to aid in the optimization.  In particular, for a given domain problem, we will explore techniques to minimize the total time to solution for a given application, such as the quantum chemical simulation mentioned above.  At this stage, we will try to answer the question of how should the compute work be distributed between the classical and quantum processors to minimize latencies and ultimately the time to solution.

An essential aspect of our existing infrastructure is the development of advanced benchmarks that test proposed hardware for performance (see Section~\ref{sec:softwareEcosystem} below). This will be a major thrust in our QC/HPC integration effort, where we will select, working with the broad user community, several key applications for testing both the quantum hardware and our integration software and workflows.  In addition, we have an existing Center for Accelerated Application Readiness (CAAR) program in which a range of application codes are examined for initial performance on new hardware and then further optimized to fully exploit the leadership class supercomputer.  The same process will be followed on both quantum simulators and newly installed hardware. This overall effort aimed at assessing hardware options, developing benchmarks, etc., overlaps with a recent DARPA initiative, Underexplored Systems for Utility-Scale Quantum Computing (US2QC), that is funding some of our work.  

The goal of this effort is to develop a fully functional facility environment for basic research in QC following a testbed model. A proposed list of activities for the work in these overlapping areas is as follows:

\begin{compactitem}
    \item Architect a flexible software infrastructure to pair HPC with a quantum simulator backend.
    \item {\color{black}Conduct a comprehensive analysis of simulators to identify candidates for large-scale integration.}  
    \item Develop and maintain vendor engagements throughout the project.
    \item Develop a prototype of the infrastructure with a quantum simulator backend, launch an alpha version, and gather user feedback.
    \item Expand infrastructure to integrate with quantum hardware located onsite or off-premises.
    \item Continue gathering user feedback and enhance the software ecosystem (compilers, libraries, benchmarking).
    \item Release the fully polished platform. Develop code optimization process following the CAAR HPC model. 
    \item Initiate alternatives analysis by plugging into the ecosystem, running benchmarks, and comparing results.
    \item Evaluate procurement options and prepare the facility for potential quantum hardware installations.
    \item Begin the remaining processes for system installation and acceptance as appropriate.
\end{compactitem}
A more detailed discussion of these efforts is presented below.

\section{HPC and QC Hardware}
\label{sec:HPC-QC-hardware}

HPC systems are known for being able to tackle a broad range of scientific problems, especially related to large-scale simulations, data analytics, and complex mathematical computations. Quantum computers, on the other hand, excel at solving specific problems using quantum algorithms that can outperform classical algorithms. Therefore, integrating both technologies is necessary to harness their strengths and create a hybrid computing infrastructure well-suited for a broader range of applications.
This holds particularly true in the NISQ era, where HPC systems can be used to simulate and verify quantum algorithms before running them on actual quantum hardware, reducing the time and cost associated with the development process. 
In this section, we discuss the various computational and quantum technologies made available at the OLCF.

\subsection{Frontier, Summit, and Advanced Computing Ecosystem}
\label{ssec:FrontierSummet}

ORNL's Frontier supercomputer holds the top spot as the world's fastest on the TOP500 list~\cite{top500}, achieving 1.192 exaflops of performance.  Frontier is the first to break the exascale barrier. The theoretical peak performance of 2 exaflops provides a tenfold increase over its predecessor, the Summit supercomputer. This system is comprised of 74 HPE Cray EX cabinets containing over 9,408 AMD-powered nodes and 37,000 GPUs interconnected by a Slingshot Dragonfly Network with 270 TB/s of bisection bandwidth. Designed to tackle the most pressing challenges in energy, economics, and national security, Frontier enables scientists to pioneer technologies crucial for the nation's future. Its architecture features the 3rd Gen EPYC processors and AMD Instinct MI250X GPU accelerators.

Meanwhile, Summit, ORNL's previous flagship supercomputer, is still operating at the OLCF. With a capability of 200 petaflops, Summit currently holds the 9th spot in the TOP500. The system is powered by IBM POWER9 CPUs, Nvidia V100 GPUs, and uses a non-blocking fat-tree topology built on Mellanox EDR InfiniBand.

Finally, the Advanced Computing Ecosystem (ACE) testbed, a distinctive capability of OLCF, offers a centralized sandbox for deploying diverse computing and data resources. It enables the evaluation of various workloads across different system architectures, fostering the development of new HPC technologies relevant to OLCF and DOE missions. This open-access environment comprises HPC production-capable resources, empowering researchers and system architects to explore existing and emerging technologies without the constraints of a production environment.

\subsection{Quantum Hardware}

Quantum hardware has made significant strides, although it remains in an early stage compared to classical computing. Various architectures are currently being explored, such as superconducting qubits~\cite{superconducting_2019,superconducting_2020}, trapped ions~\cite{trapped_ions_2019}, silicon spin qubits~\cite{Quantum_Dots_RMP_2023}, photonic qubits~\cite{photonic_review_short_2019,photonic_review_long_2018},  nitrogen vacancy centers in diamond~\cite{NV_2021}, and neutral atom qubits~\cite{Neutral_2020} or topological qubits~\cite{Topological_2008,Topological_2022}. Each of these architectures offers unique features and capabilities, making it difficult to predict which technology will best meet DiVincenzo criteria~\cite{divincenzo_dp_2000} for the physical implementation of QC.

Through the QCUP, users have access to premium QC devices with superconducting and trapped-ion qubit implementations, enabling them to explore the benefits of different architectures and helping them decide which one is better suited to address their complex computational challenges.

Superconducting qubits are comprised of electrical circuits operated at low (superconducting) temperatures. Their similarity with traditional solid-state integrated circuit technologies simplifies their design, fabrication, and even scaling approaches~\cite{superconducting_2019, superconducting_2020}. 
The short coherence times of these systems, on the order of a few hundred microseconds, are compensated by their high controllability which allows for two-qubit operations to be realized on timescales of only a couple hundred nanosecods. 
In currently available devices, however, only nearest neighbor connectivity is typically supported, which can often increase the number of operations required for entangling distant qubits. 
QCUP offers access to IBM Quantum~\cite{ibmquantum} and Rigetti~\cite{rigetti} superconducting platforms. Both vendors allow for (optional) pulse-level control of their hardware, enabling precise manipulation of the qubits that could lead to improved operation fidelity and reduced error rates, enhancing the overall performance of QAs. Achieving optimal results with pulse control, however, requires sophisticated calibration and tuning procedures for each controlled qubit, which can pose challenges for users. In addition to this feature, IBM Quantum offers mid-circuit measurement support, allowing for more flexibility in algorithm design. 

Trapped-ion qubits are charged atomic particles confined by electromagnetic fields. They exhibit coherence times on the order of minutes, and two-qubit gates that can be realized on timescales of a few hundred microseconds~\cite{trapped_ions_2019}.
Unlike other implementations, trapped-ion qubits are fundamentally identical, somewhat simplifying their control and related calibration procedures. Furthermore, these systems offer all-to-all connectivity, facilitating easier direct entanglement between distant pairs of qubits. Despite this, realizing necessary control while scaling up to larger qubit-size devices is still difficult. Quantinuum~\cite{quantinuum} and IonQ~\cite{ionq} trapped-ion platforms can be accessed through QCUP. These systems prioritize the use of native parameterized two-qubit gates, which can simplify circuit design and algorithmic expressiveness, while reducing their runtime resource requirements.

Numerous other vendors are developing concepts and hardware for quantum computing applications in a rapidly evolving technology landscape.  Two examples are Quantum Brilliance (QB)~\cite{qb} that employs nitrogen vacancy centers in diamond at ambient temperatures and IQM Quantum Computers~\cite{iqm} that utilizes superconducting qubits. 

\subsection{Hardware Integration}

\begin{figure}
\includegraphics[scale=0.113]{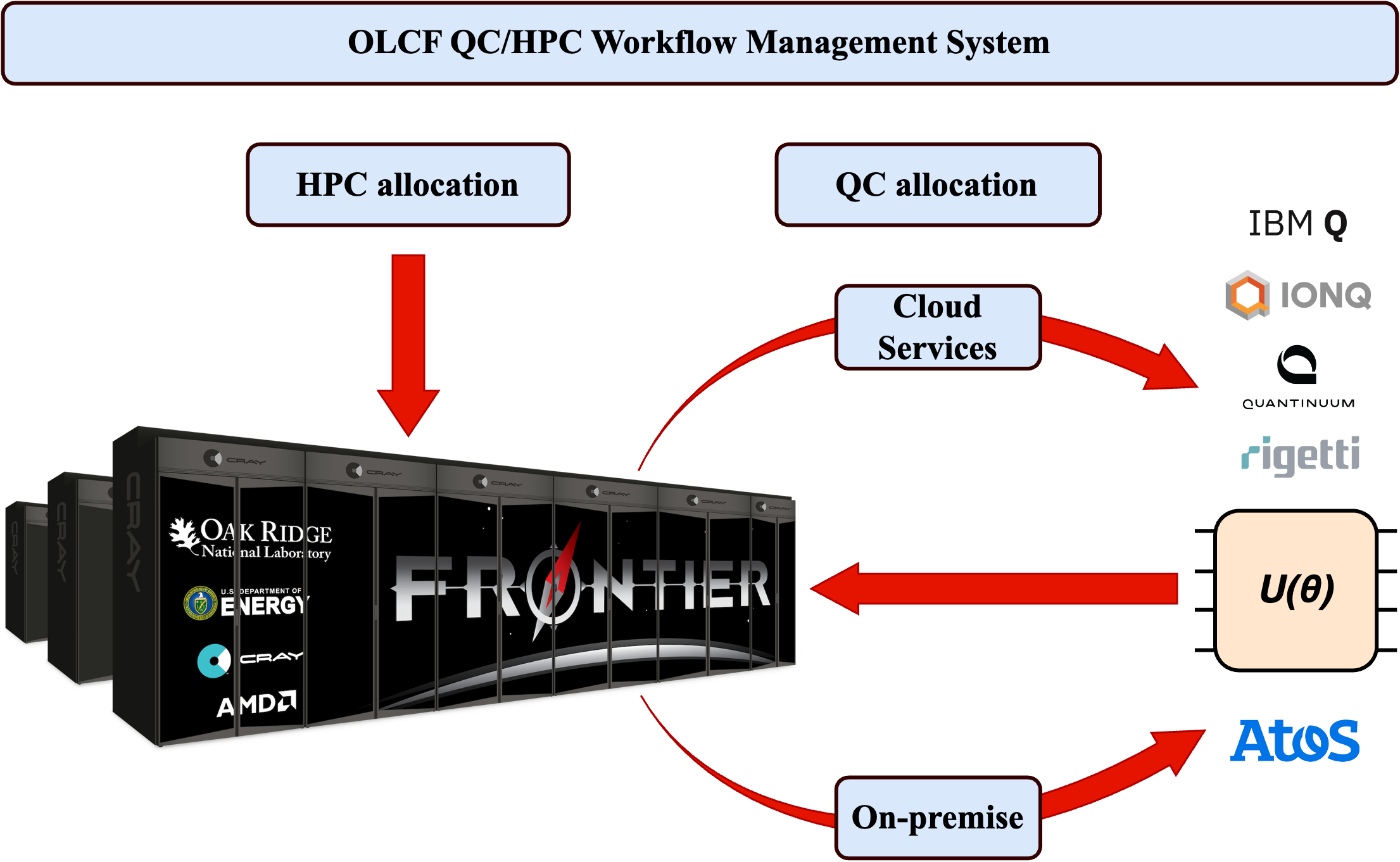}
\caption{Schematic of the QC/HPC integration at OLCF. This is a high-level view of the ideal state of the framework presented in Section~\ref{sec:integrationFramework}. The framework will integrate simulators, cloud-based quantum devices, and on-premises quantum hardware in a seamless way. 
}
\label{fig:olcf-hpc+qc}
\end{figure}

As mentioned in the previous section, various types of QC architectures exist to examine as integration possibilities. At this stage, however, it is unclear which one provides the performance and scalability desired for an on-premise QC/HPC system. Therefore, assessment and benchmarking with various QC/HPC integration prototypes are key objectives of this project. 
{\color{black} Successful integration of large-scale superconducting or ion trap quantum hardware with HPC systems hinges on overcoming several critical challenges. These include achieving high-fidelity qubit control to prevent decoherence, implementing robust error correction mechanisms, and ensuring low-latency communication interfaces between quantum and classical processors. Managing heat dissipation at cryogenic temperatures and maintaining ultra-high vacuum conditions are also pivotal. Additionally, scalability is constrained by the complexity of interconnecting numerous qubits and minimizing noise. Addressing these physical limits is essential for achieving coherent and scalable quantum operations integrated with high-performance classical computing, enabling advanced scientific and computational applications.}
Figure~\ref{fig:olcf-hpc+qc} depicts a schematic of the future QC/HPC integration at OLCF. 
For a detailed discussion of the software aspect of this effort, see Section~\ref{sec:integrationFramework}.

\subsection{Interfaces and Connectivity} 

There are two possible motifs for QPU integration into an HPC system, loose and tight. To date, the only realistic option has been loose integration in which the quantum device is physically detached from the HPC system, but connected by a network. The QPU can be located either on or off premises.  In the tight integration model, the QPU is located directly on a node in close proximity to the CPU and/or GPU hardware. This latter approach allows for all three components (CPU, GPU, and QPU) to be controlled by the same coherent software/workflow system. For the most part, here, we assume the loose integration model that is practical today, and in the most immediate future, but we note that our framework discussed in Section~\ref{sec:integrationFramework}, encapsulates the possibility of a tight integration model as well. 

{\color{black}
In the loose integration models, where QC systems are separate from HPC systems, communication occurs via network interfaces such as Infiniband or CXL. The network facilitates the exchange of requests between the HPC host and the QC hardware. In both co-located and remote server models, communication latency can be reduced by optimizing network protocols for efficient data transfer and employing advanced error correction techniques. Additionally, integrating quantum-specific middleware and optimizing software algorithms can streamline communication between HPC and QC infrastructure, ensuring minimal delay~\cite{qchpc, 7761628}. 
}

Cloud-based quantum accelerators pose latency challenges when coupled with resources located at the OLCF. We are investigating techniques for reducing latency to these devices by using available scientific networks such as ESNet and Internet2, whenever possible.
Future optimization will take into account route quality to make the most suitable choice between ESNet and Intranet2 in order to minimize communication latency and maximise network performance.

{\color{black}
Although HPC/QC interfaces rely on purely classical interfaces, the rapid development of quantum networks and their anticipated integration with HPC/QC systems are worth mentioning. Quantum networks play a crucial role in advancing quantum information science (QIS), enabling key applications such as distributed quantum computing, secure communications, and the development of a quantum internet by connecting quantum nodes such as quantum processors, quantum memories and repeaters, and quantum sensors~\cite{wehner2018quantum,azuma2023quantum}. In quantum computing, the computational power scales exponentially with the number of coherent qubits, underscoring the preference for a unified quantum computing system rather than fragmented smaller units. In this context, it is highly desirable to network quantum processors via quantum channels, first locally using quantum interconnects and quantum transducers to accommodate different qubit platforms operating over different energy scales and degrees of freedom, then over longer distances using quantum repeaters~\cite{wehner2018quantum,lauk2020perspectives,awschalom2021development}. Within this framework, future quantum networks of distributed quantum processors will need to be integrated with HPC/QC systems and offer functionalities such as implementing entanglement distribution protocols, teleportation, etc. Some recent ORNL successes towards this long-term goal are the development of a Quantum Local Area Network (QLAN)~\cite{AlshowkanPRXQ2021} to enhance entanglement distribution across existing fiber networks, as well as the demonstration of quantum and classical signals coexisting in the same optical fiber infrastructure~\cite{PeachLuPRAp2024}.
}

\section{User Base}
\label{sec:userbase}

The QCUP enables access to diverse quantum resources and introduces a wide range of users to OLCF's computing ecosystem. A major aim of our effort is to facilitate a smooth transition of our HPC community to quantum resources as they become an essential part of the computational science toolbox. Through this program, users can explore new computational research applications and potentially accelerate existing ones using QPUs. Existing QCUP projects include studies related to advanced scientific computing, high-energy physics, fusion energy science, and more.
QCUP projects are assigned OLCF Scientific Liaisons, experts in scientific domains and computation, who guide users on using QCUP resources effectively.

Analyzing QCUP's growth over time showcases the community's desire for quantum resources and the need for them going forward. As of the end of 2023, QCUP supports 80 total projects and 271 users, compared to 52 projects and 117 users in 2020~\cite{OA2023, OA2020}. Therefore, since the emergence of QCUP in OLCF, the total number of projects has increased by $\sim$54\%, while the total number of active users increased by $\sim$132\%. The growing interest in QC and the higher demand for quantum resources has led to an increase in the number of QCUP users creating new projects, or being added to existing ones.

{\color{black} QCUP users and their project activities have provided insights into the readiness of commercial quantum computing systems for scientific computing applications. This progress is tracked through online and self-reported peer-reviewed publications, which provide a technical evaluation of results obtained from using these remotely accessed systems. More than 200 publications have been compiled to date across a similar number of projects. Over the course of the program, reported publications have emphasized exploratory evaluations of new quantum algorithms with concrete examples being the testing of variational, machine learning, combinatorial optimization, and modeling and simulation methods among several others. Notably, these experimental evaluations have required interactions between the user host system for pre- and post-processing the input/output of the quantum program, but they lack integrated interactions with the QPU due to the remote access model. This motivates the considerations here for what new algorithmic methods could be enabled by tighter integration between the host and QPU systems.

The lack of tighter integration has been identified in the QCUP program as a significant barrier for the continued development of novel algorithmic methods that require near-real time processing of measurement-based conditional programming and noise characterization of the hardware itself. Presently, this limitation is due primarily to the lack of methods for programming the remote hardware controls system, i.e., commercial systems significantly restrict the available conventional computing resources and operations for user programming. This is part of the larger system concern of uncoordinated management of computational resources across multiple users and administrators. The integration of hardware control and system administration will be essential for developing performance-optimized applications that use both conventional and quantum computing resources. For example, few remote resources permit direct connections to QPU resources, instead relying on intermediate servers and queues to manage the serial nature of multi-user interactions.  Consequently, actual observed latencies  range from hours to days depending on system administration, queuing policy, and demand. This is in addition to the typical networking latencies, ranging up to a few seconds, for communication between the host and remote systems.} 

\subsection{User Support for Cross-domain Integration}

To help establish a pipeline between HPC and QCUP users, the former are provided libraries and software supported by QCUP's vendors on select OLCF systems. Allowing HPC users to access the various QC software options enables that community to test quantum workloads in their applications. On the other end of the spectrum, QCUP users are introduced to the benefits that an HPC environment can provide to the various QC libraries that are typically used locally or via the cloud. The users have the opportunity to access HPC platforms through OLCF's training events and specific QC/HPC allocation projects. Targeting both user groups in this manner leads to overall growth and diversity amongst both user bases. We are working closely with the QCUP vendors on establishing the infrastructure needed to further support hybrid QC/HPC computations at OLCF. By collaborating in software development within a given vendor's quantum library, we are able to guide how their API may interface with an HPC environment for users.

\subsection{User Engagement and Feedback}

As HPC and QC each require specialized domain knowledge and skills, providing necessary training and support to users helps them operate both individual and cross-domain computing resources effectively. Proactively integrating user feedback aligns operations with user needs and expectations, helps identify bugs, and improves system usage efficiency. User insights reveal the actual use of quantum and HPC resources compared to initial assumptions. Feedback also highlights potential accessibility or security concerns and pinpoints which areas training and educational resources are most necessary or lacking. Considering the rapid advancements in quantum-enhanced supercomputing, regularly and promptly incorporating user feedback is crucial for directing future developments. A systematic method to gather feedback includes ongoing forums, surveys, and support tickets to fully understand the use cases and user experience.

\subsection{QC/HPC Workforce Development}

In order to leverage the potential of hybrid QC/HPC, there is a critical need to train a wide array of HPC users in QC methods. This includes not only quantum physicists but also software developers, data scientists, and industry specialists. Expanding cross-domain training programs will empower users to apply quantum computing advantages in fields like climate science, AI, materials science, and drug discovery---enhancing global technological competitiveness through dedicated training of staff and up-to-date resources.

\section{Science Applications} 
\label{sec:applications}

Integrating QC into an HPC environment offers new computational capabilities with the potential to advance domain sciences relevant to the DOE's mission, and the wider scientific progress, beyond the reach of classical computing alone. The greatest benefit in the NISQ era would come from the integration of quantum hardware into HPC workflows, where portions of the problems that are hard to solve on a classical computer, but are still manageable by noisy quantum devices with a small number of error-corrected qubits, would be handled by the quantum component of the overall system. {\color{black}In what follows, we discuss a few applications encompassing potential science drivers that, based on previous work, will benefit from the integration of QC into an HPC environment.}

\subsection{Quantum Many-Body Dynamics}

The first class of problems suited to a QC/HPC workflow is based on
Feynman's original vision of QC as a quantum simulator for inherently quantum-mechanical systems. 
This includes a wide range of problems in the physical and biological sciences that can be addressed by embedding a quantum solver, such as the Hamiltonian imaginary time evolution~\cite{motta2020determining} or the Variational Quantum Eigensolver (VQE)~\cite{cerezo2022variational}, into a coarse grained or classical simulation to provide quantum accuracy to active centers in molecular simulation~\cite{Ma2020} or the fermionic correlations in the description of correlated materials. Thus, the incorporation of these QC algorithms into embedded approaches, such as the impurity solvers in Dynamical Mean-Field Theory~\cite{Bauer2016,Backes2023}, the Dynamic Cluster Approximation, or the time evolution step in Quantum Monte Carlo (QMC)~\cite{Huggins2022,kanno2023quantum}, will alleviate the sign problem these methods suffer from on classical computers. The QC algorithms will also enable higher fidelity calculations to guide the computational discovery of materials for quantum sensing, magnetic and battery applications, QMC based reactions in Molecular Dynamics simulations for catalysis and active sites in biological molecules for bio-energy and drug discovery.
{\color{black}
The framework presented here will enable the tighter coupling of QC algorithms such as these and classical steps to enable the approximate or mean-field treatment of extended systems on classical HPC systems by embedding methods. Thus the coupled QC/HPC workflow will be able to take advantage of the efficient solution of the quantum mechanical many-body problem on the quantum device and the computational flexibility of the classical HPC system.} 

\subsection{Continuum Mechanics Simulations}

A quantum linear solver algorithm (QLSA) for sparse systems has the potential to accelerate the solution of partial differential equations (PDEs) for continuum mechanics simulations like fluid dynamics~\cite{gaitan_f_2020} and heat transfer~\cite{wei_sj_2023}. There have been many use cases demonstrating the use of QLSAs, such as the Harrow-Hassidim-Llyod algorithm~\cite{harrow2009quantum} and the variational quantum linear solver~\cite{bravo2023variational}, for solving  fluid flow problems. The problems include both ideal fluid flow problems or linearized versions of the Navier--Stokes equations---nonlinear PDEs that govern the evolution of fluid flows. Of particular interest to solve practical nonlinear simulations is the development of hybrid quantum-classical solvers, and there has been preliminary success in the fluid dynamics community. Some of these nonlinear solvers relied on QC for intense computations, such as chemical kinetics~\cite{becerra2022quantum}, while others used iterative methods to switch between HPC and QC resources~\cite{bharadwaj2023hybrid}. While current QC hardware sizes are far from being competitive with classical computers for complex engineering problems, offloading independent components of a nonlinear solver to a quantum processor can help simulate complex problems in the near term. {\color{black}Our team has applied the Harrow-Hassidim-Llyod algorithm to solve an idealized fluid flow problem~\cite{Meena2024}. The current framework will enable the integration of such quantum algorithms into more practical fluid flow (and other continuum mechanics) simulations, where a particular module of the simulation can be seamlessly offloaded to the quantum computer.}

\subsection{Quantum-Enhanced Machine Learning}

The field of Quantum Machine Learning (QML) combines ML algorithms with QC, aiming to enable models with some practical advantage over those implemented solely on classical hardware~\cite{biamontej2017}. In the current NISQ era, QML has focused on the development of quantum-enhanced ML models that, at some point along their realization, encode classical data into a quantum computer and process it by means of parametrized quantum circuits, in the hope that these quantum layers can recognize and learn patterns that are difficult for a classical processor to produce. This approach has shown some success in the form of, for example, quantum generative models such as Quantum Born machines, which outperform state-of-the-art classical generative models when trained on small datasets~\cite{hibatallahm2024, houw2023}, a highly desirable feature for multiple real-life applications, for which available data is scarce. {\color{black}In previous work, we have used classical simulations to compare the classification performance of QML models trained on small synthetic cancer pathology reports, to that of classical ones~\cite{hamilton2023}. Full integration with HPC resources could enable similar studies for QML models trained on larger datasets, and hybrid models in which the large language models pre-processing real reports are fine tuned during training.}

\subsection{Quantum Optimization}

Conventional optimization methods such as Bayesian optimization~\cite{doi.org/10.1002/adom.202002226}, genetic algorithms~\cite{doi.org/10.1021/acscentsci.8b00802}, and needle optimization~\cite{doi.org/10.1038/nature13883}, struggle to locate global optimal solutions within discrete search spaces, often converging to local minima instead~\cite{10.1021/acsenergylett.2c01969}. Additionally, assessing surrogate models on classical processors post model training becomes problematic when dealing with large search spaces. QC presents a promising avenue for addressing these challenges with QAs that have demonstrated advantage in combinatorial optimization tasks. These algorithms provide acceleration to the optimization process~\cite{GSLBrandao2022fasterquantum, Gilliam2021groveradaptive}, 
particularly in large design spaces. As an example, quantum-enhanced active learning algorithms~\cite{arkopal_d_2023} offer a new approach to optimization problems. This approach enables the identification of optimal structures surpassing the performance attainable through classical optimization methods on classical computers. 
{\color{black}In previous work, we have analyzed the performance of active learning algorithms enhanced with quantum approximate optimization algorithm subroutines for metamaterial design~\cite{kim.suh2024}. The results from that study suggest considerable speedups when solving this optimization problem on hybrid QC/HPC systems.}

\section{Software Ecosystem}
\label{sec:softwareEcosystem}

Although building hardware is central to achieving practical QC, another crucial component that must be developed in parallel is a fully operational software stack. It will contain packages ranging from low-level hardware control tools, through mid-layer programs for circuit processing and transpilation, to high-level algorithmic building blocks. As already outlined, at least some parts of the quantum workflows will rely on (potentially involved) classical computations, further highlighting the importance of their early integration into large-scale HPC centers. 

The level of interplay between classical and quantum hardware will vary dramatically between different use cases. This is something the software stack will have to facilitate. Some of the classically challenging computations (e.g., decoding the error syndrome, that is operations used to determine how errors are to be corrected during a quantum error correction process) will require low-latency, almost continuous interactions between the quantum and classical computers, while others (e.g., transpilation -- the mapping of high-level QAs onto specific hardware architectures) will be able to be performed on the classical hardware, largely independently. Furthermore, before fully error-corrected quantum hardware becomes available, simulators 
will continue to play an important role in exploring smaller-scale QAs, finding ways to mitigate the effects of noise, and studying more effective control techniques. Thus, they must be accounted for in any QC/HPC integration efforts. 

In Section~\ref{sec:integrationFramework}, we present a comprehensive discussion related to a software infrastructure framework for integrating quantum devices within an HPC center such as OLCF. First, however, we provide a brief overview of the current state of quantum software frameworks and tools, discuss ongoing efforts to quantify the performance of quantum (and accompanying classical) devices, and highlight some of the key simulation techniques and packages.

\subsection{Application Frameworks and Tools}
\label{ssec:applicationFrameworks}

Much of the recent software development efforts aimed at operating and using the currently available devices have been mainly led by various vendors (e.g., IBM, Google, Quantinuum, Rigetti). Each vendor has developed their own software frameworks (such as Qiskit~\cite{qiskit}, Cirq~\cite{cirqpackage}, Tket~\cite{sivarajah2020tket}, PyQuil~\cite{computing2019pyquil}, respectively) for accessing their specific hardware, and more general circuit and algorithm design and processing. Some companies with limited or no available hardware have also managed to build tools that have caught the attention of the broader community (e.g., Xanadu with the Pennylane~\cite{bergholm2018pennylane} framework). 

Large portions---typically the high-level components---of these packages have been open-sourced, providing a considerable benefit to researchers and students alike. However, to further differentiate themselves from other companies in the future, vendors could possibly keep larger and larger parts of their software ecosystems proprietary. Furthermore, support for HPC systems is currently often limited to CPU-only, or single-GPU code that runs in an embarrassingly parallel mode, without fully taking advantage of the highly distributed, multi GPU-based hardware that dominates large supercomputing centers such as the OLCF.
 
Hence, further effort will have to be made to build software tools that guarantee both sustainability and openness, but also embrace the requirements for seamless interaction between applications, quantum hardware from various vendors, as well as highly distributed, often GPU-based HPC infrastructure. 
One leading attempt related to realizing such interoperability is through the QIR Alliance~\cite{qiralliance}. This multiorganizational effort (involving ORNL, Microsoft, Nvidia, Quantinuum, Rigetti, and Qci) aims to define an abstraction standard for backend-independent specification of quantum programs~\cite{qiralliancestandard} based on the LLVM~\cite{llvm} toolchain. The effort also aims to implement this specification through ready-to-use tools~\cite{qirallianceprojects} that can be utilized by application as well as hardware and various simulator developers. 

An existing example of such a tool is ORNL's XACC~\cite{mccaskey2020xacc}, an extensible compilation framework for hybrid quantum-classical computing architectures that has been built with support for large-scale HPC-based classical compute in mind. This has been done in a quantum hardware-agnostic way able to adhere to the QIR specification mentioned above~\cite{wong2023qir}: XACC allows interfacing with machines from various vendors. 
Other efforts to establish portable representations of quantum programs also exist, with arguably the IBM-led OpenQASM initiative~\cite{cross2017open,cross2022openqasm} being the most popular and widely supported by a variety of quantum hardware vendors and simulator developers. It provides an easily human-readable program specification language in terms of quantum gates, measurements, and conditionals. More recent extensions also allow for lower-level control pulse definitions. Although it is too early to see what particular proposal will become the standard in the long term, it is crucial for our community to strive towards open and widely accessible approaches. ORNL's long-standing institutional know-how and experience could play an important role in helping to lead this effort.

\subsection{Benchmarking}
\label{ssec:bechmarking}

Given we are in the early stages of quantum hardware development, no universally accepted quantum-system benchmarks 
have been fully established as the industry standard.
Arguably, one of the most widely used approaches is related to the IBM-proposed Quantum Volume (QV)~\cite{bishop2017quantum} metric, which along with the circuit-layer operations per second (CLOPS) measure~\cite{wack2021quality}, aims to capture the hardware performance in terms of qubit quality, system scaling and speed of operations.
At its core, IBM's proposed method is largely based on executing random circuits, with the purpose of capturing general performance characteristics, and not ones too closely tied to a particular application type.
Other benchmarking suites, however, that rely more closely on exploring a subset of specific applications or circuits that users might be directly interested in running, also exist (e.g., qasmBENCH~\cite{li2023qasmbench}, SupermarQ~\cite{tomesh2022supermarq} or quantum LINPACK~\cite{dong2021random}). 

Another recent benchmark for assessing the effectiveness and practicality of a quantum advantage has been introduced based on the term \textit{quantum utility}~\cite{herrmann2023quantum}, where various aspects of assessing the performance of applications and algorithms are considered. These include scalability, compilability, connectivity, robustness, and parallelizability. Similar to the international standard of Technology Readiness Levels~\cite{mankins1995technology}, the authors introduce the concept of Application Readiness Levels to monitor the readiness of quantum applications to achieve quantum advantage and, eventually, quantum utility.
Finally, some vendors (or laboratories) simply stick to metrics that capture gate fidelities along with ratios of gate to coherence times. Although these are generally not seen as adequate for capturing algorithmic performance (or integration with any classical hardware), they can still be useful as a first-pass comparison between the capabilities of different quantum devices. 

More work is needed to establish optimal ways to describe and properly compare the performance of different quantum (and the accompanying classical) systems. 
These metrics have to encapsulate features often not present in more traditional classical-only computing infrastructure, such as qubit connectivity, topology, or the variability in the native gate sets. 
ORNL has a long history of developing benchmarking technologies (e.g., LINPACK~\cite{dongarra2003linpack}, SHOC~\cite{danalis2010scalable}) and using them for exhaustive comparisons and performance studies~\cite{joubert2012analysis,zimmer2019evaluation,atchley2023frontier,malaya2023experiences} of the evolving HPC infrastructure. This, together with our growing expertise in quantum sciences will help us to closely work with the broader scientific community to help build the right tools and help establish standards in approaches to QC/HPC benchmarking.

\subsection{Simulators}
\label{ssec:simulators}

\begin{figure}
\includegraphics[scale=0.35]{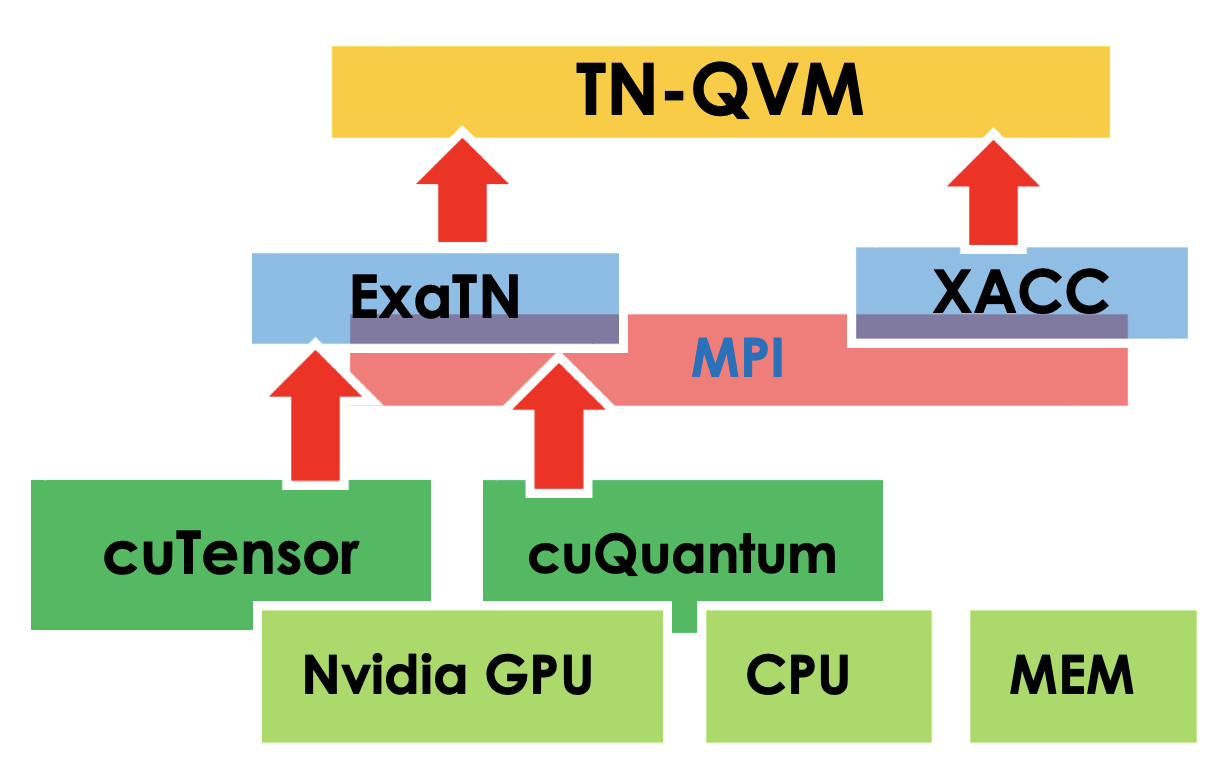}
\caption{Schematic showing how the TN-QVM, an ORNL-developed tensor-network based accelerator, integrates with various other software packages to form a powerful tool for simulating large circuits and algorithms on HPC platforms.
}
\label{fig:tnqvm}
\end{figure}

Classical simulators allow for accessible development, testing, and analysis of quantum algorithms~\cite{xu2023herculean}. They also help with exploration of general quantum system evolution, all while providing a controlled environment for investigating the effects of noise and resulting errors. Although quantum computers hold an immense promise for the future, classical simulators remain a vital component in the QC toolkit, enabling progress and innovation in the field.

At ORNL, we deploy a number of simulators and related tools, able to utilize the capabilities of our HPC infrastructure (see Section~\ref{ssec:FrontierSummet}). These include TN-QVM~\cite{Nguyen2021TensorNQ, 10313902}, an ORNL-developed tensor-network based accelerator that works together with XACC (see Section~\ref{ssec:applicationFrameworks}) and tensor-network simulator backends such as ITensor~\cite{ITensor} or ExaTN~\cite{lyakh2022exatn} for high-performance exploration of circuit evolution. Figure~\ref{fig:tnqvm} shows a schematic of how TN-QVM integrates into a larger software ecosystem. We also directly support NWQ-Sim~\cite{li2020density, li2021svsim}, a 
state vector and density matrix simulator, capable of exploring the evolution of both ideal as well as noisy circuits (NWQ-Sim is currently fully operational on Summit, and is being updated with support for Frontier's AMD-based GPUs)~\cite{9910084, 10.5555/3433701.3433718}. Other simulators that specialize in predicting quantum system dynamics (and not just gate-based circuit evolution) in the context of optimal control can also be utilized. These include e.g., QuTiP~\cite{johansson2012qutip}, qiskit-dynamics~\cite{puzzuoli2023qiskit}, or Quandary~\cite{gunther2021quandary} (the former two offer limited distributed-system support, although they are capable of running on GPUs, while the latter, through MPI-based capabilities, can take full advantage of our highly-distributed HPC-infrastructure). 

Finally, besides software-only simulators and related tools we discuss above, we are also exploring dedicated hardware-software based solutions, where purpose-built classical computers, together with specially tailored software tools, form a tightly integrated unit that can be utilized by our QC/HPC framework as yet another simulation backend. One example of such a system is the Atos Quantum Learning Machine (QLM)~\cite{atos}, which includes a programming platform, as well as a collection of high-performance software simulators and optimizers. We believe that this multi-pronged approach to support a wide variety of backend systems and solutions will play a crucial role in helping with the longevity, adaptiveness, openness, and extensibility of our broad integration efforts.

\section{Integration Framework}
\label{sec:integrationFramework}


Efforts to create frameworks built to integrate the growing QC capabilities into HPC environments are gaining traction. 
Access to the QPU or simulators still needs to be orchestrated by a resource management system. This system allocates and schedules the resources requested by the application. For example, current common frameworks in use within OLCF HPC systems are SLURM (Frontier) and IBM Platform LSF (Summit) as job schedulers. However, for a QC/HPC integrated hybrid system, jobs which tightly couple classical tasks with quantum tasks must be co-allocated and co-scheduled to ensure the optimal utilization of the QC/HPC system. Hence, the development of a QC aware resource management systems will be a significant aspect of system integration.

Our proposed framework design has the following attributes:

\begin{compactitem}
    \item It offers a versatile and generic environment for executing quantum tasks across diverse quantum platforms.
    \item It allows users to leverage any circuit composition framework, e.g., Pennylane or Qiskit, while standardizing on OpenQASM or QIR format for quantum program description.
    \item It uses an MPI-based mechanism to communicate with quantum platforms, allowing for seamless integration into the current HPC programming paradigm.
    \item It reserves and manages both HPC and QC resources concurrently.
\end{compactitem}

\subsection{Integration Space}

There are two predominant QC/HPC integration spaces to consider: the loose and tight integration models. The latter involves the direct integration of Quantum Processing Units (QPUs) 
into HPC nodes, akin to GPUs, which is not expected to be feasible in the near future. The loose integration model is our main focus; however, the design of the framework will not preclude the tight integration model. 

In the loose integration space, QC resources operate as distinct entities integrated into the broader HPC environment. This loose integration can be broken up into off-premises, where QC resources reside in remote cloud environments, and on-premises, which constitutes the main focus of our work. In the on-premises scenario, a QC is integrated into the HPC center. It is connected to classical HPC systems via high-bandwidth interconnects and a distributed file system. The connectivity facilitates communication between classical and quantum resources allowing for the acceleration of specific workflows.

Our framework offers hybrid applications mechanisms to use the computational power provided by HPC for both classical logic and quantum simulation. At the same time, it provides a seamless path towards transitioning to actual quantum hardware. This approach empowers researchers and practitioners to exploit the combined potential of HPC and QC for various scientific and computational tasks.

\subsection{QC/HPC Integration Usage Patterns}

Hybrid QC/HPC applications usage of quantum computing can be categorized into three patterns:

\begin{compactitem}
    \item \emph{In-Sequence Processing:} these applications require mid-circuit measurements, then based on classical processing, they modify the circuit during its execution.
    \item \emph{Single-Circuit:} these applications execute quantum circuits repeatedly to generate a distribution of measurements and a resulting statistical characterization. These circuits can be large and complex.
    \item \emph{Ensemble-Circuit:} these applications may execute multiple independent circuit instances to generate a distribution of measurements and a resulting statistical characterization. Circuit results are aggregated and post processed by classical logic.
\end{compactitem}

\subsection{QC/HPC Integration Models}

To satisfy the hybrid QC/HPC application usage patterns outlined above, the framework proposes support for a single QC resource model and per-job QC model. The single QC model supports real quantum HW and a single simulation resource like the ATOS system mentioned earlier. HPC jobs may reserve this single QC resource with classical HPC resources to satisfy the hybrid application's requirements.


In the per-job QC model, each classical HPC job is assigned a specific allocation of quantum resources. While at the current stage of QC technological evolution, this approach is not feasible with real quantum hardware and can be exploited for simulation purposes. Under this model, each hybrid QC/HPC job gets two simultaneous HPC node allocations: One for running the classical logic of the hybrid application and another dedicated to simulating quantum circuits. Both allocations can be specified independently, allowing tailored resource assignment for different types of hybrid applications.

{\color{black}
The integration of QC/HPC involves two key layers of job scheduling and resource allocation:
(a)~The higher-level task scheduling involves the workflow management system parsing and compiling workflows. Tasks and their dependencies are evaluated for parallel execution, and eligible tasks are delivered to HPC/QC resource managers for scheduling; (b)~At the HPC/QC resource layer, job definitions are received from the workflow management system. Resource allocation systems like SLURM and PBS manage resources among users, while workflow managers strategize task granularity based on user configuration.
}

\subsection{QC/HPC Integration Framework Overview}

The Quantum framework (QFw) proposed in this paper offers a solution that supports the integration models described above. While many frameworks in existence today allow the specification of different simulations or physical backends, they tend to force the user into a specific programming paradigm, e.g., CUDA-Q~\cite{cuda-q} or Qiskit. Recognizing that the field of QC is rapidly evolving and the appearance of new programming paradigms is inevitable, the QFw provides a way to mix and match any frontend circuit building tool with any backend simulation package. Both the frontend application and backend simulator can utilize large scale HPC resources. As shown in Figure~\ref{fig:hpc_qc_architecture} {\color{black} (Section~\ref{sec:introduction})}, the QFw allows users to leverage any circuit composition framework like Pennylane, Qiskit, CUDA-Q, etc. while still maintaining standardization in the form of common text-based OpenQASM or QIR formats for quantum program specification.

The QFw offers a set of concrete proposals for the modification of the Resource Management System to allow the simultaneous allocation of QC resources and HPC resources. This is a key requirement for hybrid applications. These modifications work for both quantum simulation as well as real hardware resources, allowing applications to transition from one to the other seamlessly.

In practice, applications which use the QFw in a simulated environment specify the HPC resources they require for the classical and quantum simulation independently. They can then use any circuit composition software suited to their respective use cases. The QFw provides a backend for the conversion of native quantum circuit structures into OpenQASM or QIR representations. This common quantum task format is passed to lower layers of the framework for further processing.

The Quantum Task Manager (QTM) layer may apply specific workflows, such as circuit cutting for southbound tasks and aggregation for northbound results. The Quantum Platform Manager (QPM) manages communication with the underlying platform. It receives the quantum tasks from the QTM in a standardized format and executes them through platform specific operations. The QPM provides a common API that can have platform dependent implementation. The QFw provides a common utility layer that can be used by any QPM implementation to streamline development. The QFw can handle multiple QPMs to support the simultaneous utilization of different quantum platforms.

Ultimately, we envision the QPM API as a way to create a standard quantum library, which not only specifies the APIs to execute quantum tasks but also serves as a standard API for accessing different aspects of a quantum platform, including but not limited to platform calibration and platform resource management. This plugin approach allows the integration of new quantum platforms seamlessly. It also provides a separation of concerns that allows applications to operate at a higher level of abstraction without worrying about hardware specific details. This design follows well established library designs used extensively within the HPC space, such as the MPI libraries~\cite{mpi_library} and the libfabric library~\cite{libfabric}. 

\subsection{Dynamic Simulation Environment}

\begin{figure}[!t]
    \centering
    \includegraphics[width=1\linewidth]{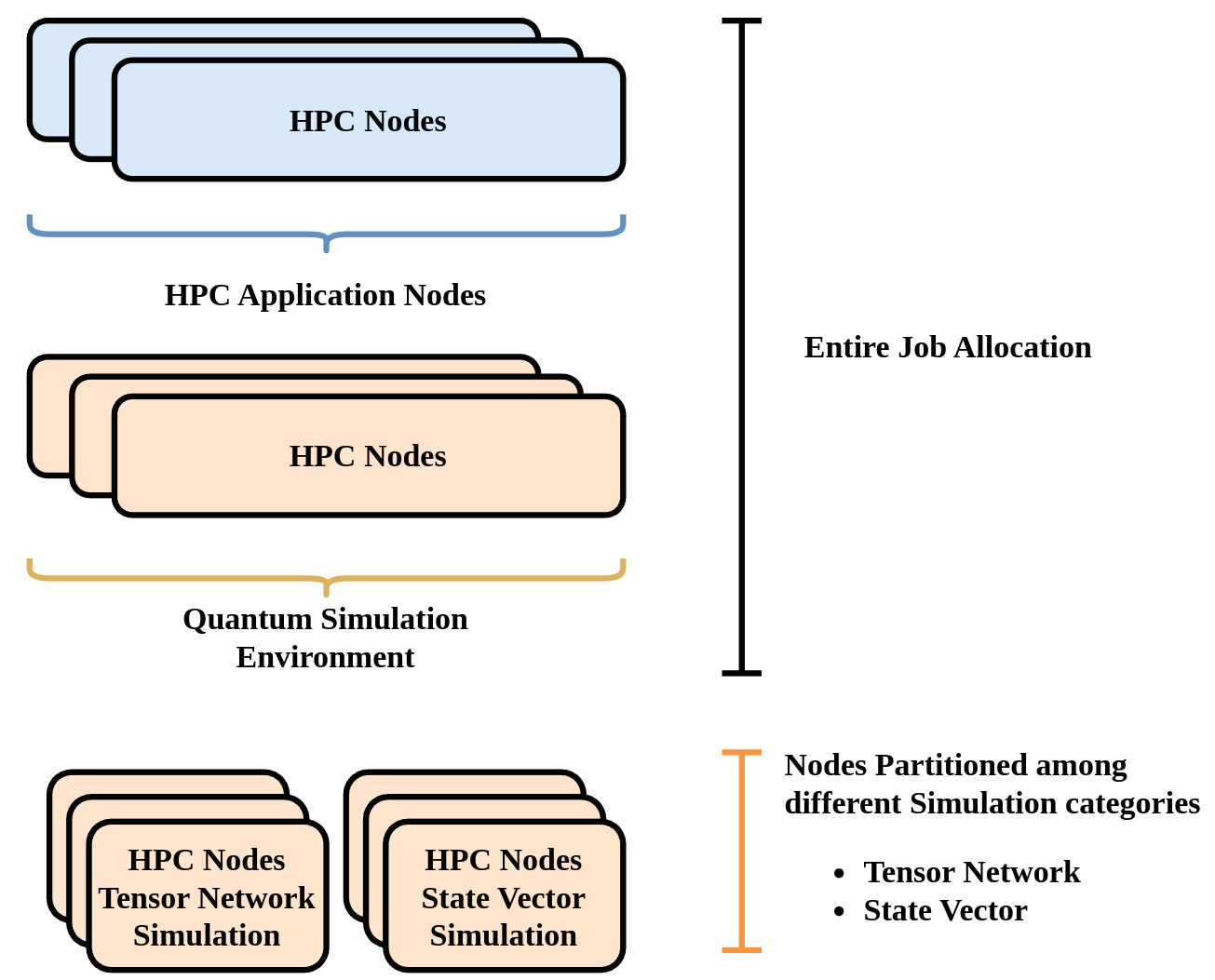}
    \caption{The dynamic simulation environment is composed of a set of classical HPC nodes. It can be partitioned into separate sets for running different types of quantum simulators. See also discussion in Sec.~\ref{ssec:simulators})}
    \label{fig:dynamic_qsim}
\end{figure}

Even after the development of fault tolerant Quantum Computers, the need for quantum simulators will persist. For the foreseeable future quantum simulators running on classical hardware will be more accessible and less expensive. It is, therefore, a logical approach to test and debug quantum programs on quantum simulators before deployment on actual hardware. To address this need, part of the QFw proposal is the development of a flexible and extensible simulation environment, depicted in Figure~\ref{fig:dynamic_qsim}. This simulation environment is capable of accommodating various simulation backends, giving the researchers the flexibility to debug and test their algorithm on the simulator that best fits their needs.

As mentioned above, the QFw offers mechanisms to allocate HPC nodes for the classical logic of the hybrid application independent of the HPC nodes used for quantum simulation. The latter set of HPC nodes are labeled the ``simulation environment''. It can be subdivided based on user input into distinct partitions responsible for executing different types of quantum simulators such as tensor network simulators and state vector simulators. When an application submits quantum tasks to the QFw for execution in the simulation environment, the framework employs heuristics—considering factors such as qubit count and gate depth—to determine the most suitable simulators for the task. User preferences supersede default heuristics, allowing for customized simulation configurations.

Upon receiving queued quantum tasks, the environment dynamically assesses the required number of simulators and configures itself accordingly to handle concurrent simulations efficiently. These simulators can operate as MPI jobs, enabling a single simulator to span multiple MPI processes and nodes. This accelerates the simulation of large quantum tasks. Conversely, for smaller quantum tasks that do not require multiple MPI processes, the environment can spawn multiple simulators on HPC resources to execute multiple quantum tasks concurrently. A combination of both run modes is also supported.

This dynamic approach optimizes the utilization of HPC resources. Moreover, it enables researchers to test their quantum algorithms across a diverse range of simulators. For instance, while some hybrid applications may benefit from tensor network simulators, others may require state vector simulators. Additionally, this approach permits applications to leverage both types of simulators simultaneously, facilitating tailored simulation strategies based on the specific needs of the hybrid QC/HPC applications.

{\color{black} A prototype of the SLURM allocation model and simulation environment components of the QFw architecture has been developed. It allocates two distinct sets of nodes: one dedicated to the hybrid application and the other to the simulation environment. A sample hybrid application example submits circuits to the simulation environment using SupermarQ to generate circuits of varying qubit numbers. The environment supports parallel simulation of these circuits and uses PRTE (Process Resource and Topology Engine) to manage resources, facilitating MPI-enabled simulators with mpirun. Currently, XACC~\cite{mccaskey2020xacc}, TN-QVM~\cite{Nguyen2021TensorNQ} and ExaTN~\cite{lyakh2022exatn} is supported as the simulation software stack backend, allowing for two levels of parallelism: independent circuit simulation and single circuit simulation. The ultimate goal is to make QFw available as an open-source project. A more thorough description of the details of the QFw will be presented in forthcoming publications that build upon this initial overview. 
}

\subsection{Workflows and End-to-End Integration}

Workflows are essential for orchestrating and managing tasks across QC and HPC systems~\cite{bieberich2023bridging, wcs2022}. The QC and HPC systems, while powerful on their own, achieve unprecedented efficiency and capability when integrated through well-defined workflows. Such workflows facilitate the structured execution of multi-stage computations that involve both environments, allowing for a coherent execution where tasks can be offloaded to the system best suited for them. For instance, a workflow might direct computationally intense data processing tasks to a classical HPC system, while quantum-specific algorithms, such as those for quantum chemistry or optimization, are routed to a QC system. This bifurcation not only maximizes the strengths of each system but also enhances overall efficiency and output quality. Future computational projects will increasingly rely on both QC and HPC systems working in tandem, with workflows managing tasks such as data streaming and signal processing between systems. This end-to-end integration is crucial for the development of applications that are scalable and adaptable to the evolving landscape of computational technologies.

\section{Summary and Outlook}
\subsection{Summary}

QC is instrumental to the future of overall scientific computing. The OLCF, through the QCUP program, has been driving toward a growing understanding of the techniques and technologies needed to make quantum HPC a reality. This paper details our efforts over the last five years in providing user access to quantum technologies and our roadmap for realizing a QC/HPC integrated environment. Through our QCUP program, we have amassed significant experience with quantum hardware, software, and integration. In this work, we have surveyed the state of the technologies that the QCUP has provided and identified several challenges we seek to overcome, starting with integrating HPC resources and quantum accelerators. 

Integration poses several challenges, ranging from network latencies to resource management, software frameworks, and programming environments.  These challenges are exacerbated by rapidly changing technology and capability. This paper presents our prototype for QFw, a framework for coupling modeling and simulation workflows with QC circuit simulations. This framework provides tools for resource management, application deployment, circuit deployment, and coupling. QFw will allow HPC application developers to explore offloading portions of HPC workflows into QC circuits in preparation for more capable technologies while allowing system software designers to understand better the orchestration of data and the necessary steps to enable software coupling through system services.

\subsection{Outlook}

As we approach quantum error correction and fault-tolerant QC that enables utility-scale quantum, it is apparent that the most effective deployments will incorporate coupled models of classical HPC and large-scale QC. The problems being solved today by integrating these systems at a small scale will inform the strategies used for more significant utility-scale investments. Future QCUP efforts will investigate on-premise quantum machines, targeting the necessary efforts to enable HPC data centers to accommodate these new technologies that are more sensitive to environmental effects. This also provides opportunities for exploring quantum networking to improve latencies of integrated components and studies examining the impacts of scaling HPC applications coupled with quantum accelerators, as scale always brings new challenges. Additionally, the OLCF will continue growing and fostering both users of quantum systems and a workforce capable of handling the challenges of these new hybrid data centers.

\section*{Acknowledgments}
This research used resources of the Oak Ridge Leadership Computing Facility at the Oak Ridge National Laboratory, which is supported by the Office of Science of the U.S. Department of Energy under Contract No. DE-AC05-00OR22725.


\end{document}